\documentstyle[prb,aps,psfig]{revtex}

\begin{document}
\title{Theory of Fast Quantum Control of Exciton Dynamics \\ in
Semiconductor Quantum Dots}
\twocolumn [\hsize\textwidth\columnwidth\hsize\csname@twocolumnfalse\endcsname
\author{C. Piermarocchi, Pochung Chen, Y. S.  Dale,  and L. J. Sham}
\address{Department of Physics, University of California San Diego, La Jolla CA
92093-0319.}

\date{\today} \maketitle

\draft

\begin{abstract}  
Optical techniques for the quantum control of the dynamics of
multiexciton states in a semiconductor quantum dot are explored in
theory.  Composite bichromatic phase-locked pulses are shown to reduce
the time of elementary quantum operations on excitons and biexcitons
by an order of magnitude or more. Analytic and numerical methods of
designing the pulse sequences are investigated.  Fidelity of the
operation is used to gauge its quality. A modified Quantum Fourier
Transform algorithm is constructed with only Rabi rotations and is
shown to reduce the number of operations. Application of the designed
pulses to the algorithm is tested by a numerical simulation.
\end{abstract} \pacs{} \date{\today}

]

\section{Introduction}

The possibility of controlling the dynamics of a quantum system has
long captured the attention of workers in a wide range of physical
systems. Quantum control can be realized by engineering a
time-dependent Hamiltonian which depends on a finite set of
parameters. In quantum chemistry, this engineering has led to the
possibility of driving chemical reactions using tailored laser pulses
and external fields.\cite{rabitz00} Quantum control has been recently
extended to semiconductor nanostructures leading, for instance, to
controlled currents,\cite{hache97} coherent control of
excitons\cite{heberle95,bonadeo} and electron spin,\cite{awschalom}
and controlled intersubband transitions of shallow donors using
terahertz radiation.\cite{cole01}

In this paper we explore optical control of the ultimate quantum
device in semiconductor nanotechnology, i.e., a quantum dot. In a
semiconductor quantum dot the electronic levels have a density of
states characteristic of a single atom. Yet, the dot is a {\it
mesoscopic} system, i.e., in contrast to the single atom case, the
quantization of the electronic levels is realized within a system that
contains actually $10^5 \sim 10^6$ atoms. A key ingredient in the
quantum control of these semiconductor nanostructures is the
robustness of the elementary excitation, the exciton. An electron-hole
pair optically excited in an undoped quantum dot feels the presence of
the large number of atoms in the material only through the static
dielectric constant and the electron and hole effective
mass.\cite{sham66} This allow us to treat excitons as excitations in
giant atoms and to control excitons with optical techniques similar to
those used for the manipulation of atoms and molecules. However,
unlike the atomic case, the dot is in a solid state environment, with
the attendant decoherence. We shall also make use of the conduction
band electron and the valence band hole as the constituents of an
exciton. A quantum dot is like an empty box that can be filled with
multiexciton complexes composed of many interacting
excitons.\cite{bayer00} In these multiexciton states, the Coulomb
correlation is taken into account and yet the spin configuration is
transparent.  The spin configuration can then be controlled by the
light polarization of the optical pulses.

The implementation of quantum algorithms is a particular case of
quantum control. The potentialities of semiconductor quantum dots in
the implementation of quantum algorithms have been readily recognized,
\cite{barenco,loss,sanders} as well as in conjunction with optical
microcavities. \cite{wang,imamoglu,sherwin} The use of optical control
of excitons in dots for quantum operations has been suggested,
\cite{troiani} and a theory of the physical implementation of quantum
algorithms in a dot using ultrafast optical pulses was investigated.
\cite{chen} Ideas for a scalable quantum computer involving excitons
in different dots and optical quantum control were
proposed.\cite{wang,biolatti,quiroga} Advances in ultrafast optics in
quantum dots make possible the manipulation of electronic excitations
in a semiconductor nanostructure with time resolution in the
femtosecond domain. So far, frequency selection is used to avoid
unwanted transitions to states out of the computational space. Laser
pulses of a narrow frequency range are too long in duration compared
with the decoherence time for quantum operations. Thus, a design of
fast control is necessary. A fast control allows us to make a
reasonable number of operations well within the decoherence time. It
may also be made an ingredient in the realization of sophisticated
error correction and decoupling schemes.\cite{viola00} We will give an
explicit design for the realization of fast control of two qubits
encoded in two antiparallel-spin excitons in a single quantum dot. The
slight increase in complexity of the optical setup is within the
capability of the current experiments.  An optimal design is an
inverse problem to the finding a state given the Hamiltonian: the
issue is to find a time-dependent optical electric field that produces
a desired result in the shortest time as possible. The required
experimental resources are realistic: lasers generating Gaussian
pulses with two different frequencies that can be phase-locked.  The
synthesis of phase locked optical pulse from separate femtosecond
lasers has been recently reported, \cite{shelton01} and here we
propose an important application of this technique.  We explore the
three different approaches to the control problem: an intuitive one
making use of the area theorem \cite{allen} and also give an
analytical tool, the cluster expansion of the evolution operation,
much used in the NMR spectroscopy, \cite{ernst87} and numerical
optimization.

The implementation of a 2-qubit Quantum Fourier Transform (QFT)
\cite{chuang00} is used as a test case for the different methods of
design. Any algorithm can be decomposed as a series of single-qubit
rotations and two-qubit conditional rotations. \cite{chuang00} A
series of two-color phase-locked optical pulses is suggested to
realize the fast control of these fundamental rotations. Fidelity
\cite{poyatos} is used to gauge the quality of the operations and of
the complete algorithm of QFT within the decoherence time. Error
correction could be added later to improve the result.

The paper is organized as follow. Section~\ref{multiexcitons} gives
the structure of the multiexciton states from a microscopic model for
a quantum dot.  Section~\ref{control} explains the principles of three
methods of design of optical pulses for a fast control in a subspace
of multiexciton states and compares their results in fundamental
quantum operations. Section~\ref{QFT} contains a numerical simulation
of the QFT algorithm in a quantum dot. The simulation takes into
account the microscopic details of the laser-exciton dynamics,
including decoherence and the presence of multiexciton levels outside
the computational space. Details of the decomposition of the QFT in
terms of only Rabi rotations for a general $n$ qubit system are
relegated to the appendix. Section~\ref{discuss} summarizes and draws
a number of conclusions. A brief description of the key idea of pulse
shaping and the application to a different quantum algorithm were
published in Ref.~\onlinecite{chen}.

\section{Multiexciton states}
\label{multiexcitons}

\begin{figure}
\centerline{\psfig{file=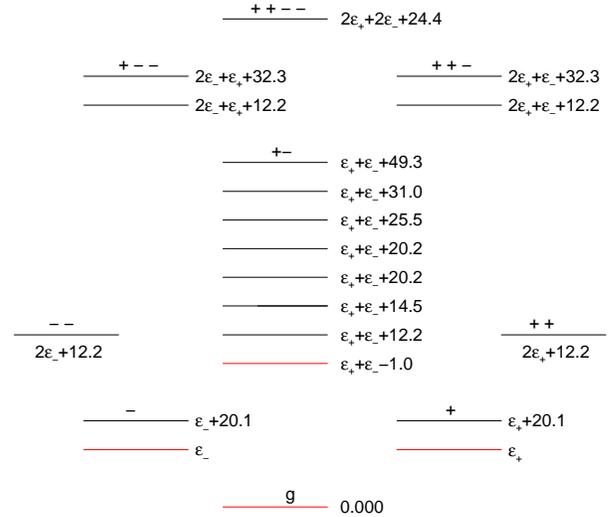,width=8. cm,angle=0}}
\vspace{0.5 cm}
\caption{Energy levels of the multiexciton states in a quantum dot in
meV.  $\varepsilon_+ = \varepsilon_-=1764$ meV.}
\label{multiex}
\end{figure}

 The energies and wave functions of the multi-exciton states in a dot
are calculated starting from two confined levels of electrons and
holes each in a parallelepiped QD.\cite{barencodup95} The electronic
levels included are the first two states deriving from the
localization of $s$-like conduction band states. They carry a spin
$\pm 1/2$. The hole levels derive from the localization of states in
the $p$-like valence band heavy holes carrying a $\pm 3/2$ total spin
in the direction of the growth axis.  The size of the dot,
40~nm$\times$35~nm$\times$5~nm, is typical of interface fluctuation
quantum dots.\cite{bonadeoprl98} Only Coulomb interaction between the
carriers which conserves their conduction or valence band indices is
taken into account exactly. This amounts to neglecting the
electron-hole exchange which gives a fine structure of the excitonic
levels depending on the symmetry of the dot. We calculated this effect
to be of the order of a few $\mu$eV, which can thus be safely
neglected in the discussion of fast control considered in this paper.
Fig.~\ref{multiex} shows the energy structure of the multiexciton
states.  The multiexciton levels include zero, one, two, three, and
four excitons in the dot. The choice of two levels each of electrons
and holes limits the resultant number of excitons to four.  The $+$ or
$-$ refers to the polarization of the light that has to be used to
create each exciton.  Only optically active multiexciton states are
shown. Since the optically forbidden multiexciton states are not the
source of unintended dynamics, they are removed from the following
discussion.  Our model adopts the measured dipole moment of $75$ Debye
for the single exciton in a single GaAs fluctuation dot
\cite{stievater01} and the transition matrix elements between the
multiexciton states are then calculated. The later values are used in
the numerical simulations in section~\ref{QFT}. Note that the values
of the dipole moments in this kind of systems are one or two orders of
magnitude higher than those of atoms.  Theoretical estimates suggest
that this giant dipole effect seems to be stronger in quantum dots
generated by monolayer fluctuations than the self-assembled
dots.\cite{andreani}

\section{Quantum control of multiexciton states}
\label{control}

\subsection{Formulation of the Problem}

Quantum control consists in designing a time-dependent addition to the
system Hamiltonian which will drive the state of the system from a
prepared state to a designated state within a number of desirable
conditions. In this paper, we shall focus on the dynamics of two
excitons in a quantum dot. The controlling agent is a sequence of
laser pulses. The dynamics of this two-exciton system not only serves
as a powerful illustration of the more general case of multiple
discrete states but also to form a set of ``universal gates'', i.e.,
fundamental operations in terms of which any quantum computation may
be constructed.\cite{chuang00} The system is not closed. In addition
to the laser interaction, the quantum dot has other excitonic states
and its environment of the substrate and other dots is a source of
decoherence. The time limitation due to decoherence and the resonance
conditions to avoid the unintended dynamics form two contradictory
requirements under which the dynamics of the two excitons in a dot
must be optimized.

For the system of two excitons, we use a basis set of four states, in
order, the ground state $|0\rangle$, the two excitons with the lowest
energy at 1764 meV above the ground state and opposite polarizations
denoted by $|+\rangle$ and $|-\rangle$ and the biexciton state
$|-+\rangle$ at 3527 meV.  A $\sigma+$ polarized photon can drive the
excitonic transition $|0\rangle \rightarrow |+\rangle$ and the
biexcitonic transition $|-\rangle \rightarrow |-+\rangle$.  The two
transitions have different energies due to the Coulomb binding energy
of the biexciton. We can write the Hamiltonian of the four-level
systems coupled to an external electromagnetic field with $\sigma^+$
polarization, treated classically, in the form
\begin{equation} 
H^+=\left[\begin{array}{cccc} 
0& \Omega_+(t)/2 & 0 & 0 \\
\Omega^*_+(t)/2& \epsilon_+& 0 & 0 \\ 
0& 0 & \epsilon_- & f \Omega_+(t)/2 \\ 
0& 0 & f \Omega^*_+(t)/2 & \epsilon_{-+}
\end{array}\right]~,
\end{equation} 
where $ \Omega_+(t) =\sum_j d_+E_{+,j}(t-\tau_j)$ represents a time
dependent Rabi energy provided by a train of phase-locked optical
pulses. The dipole moment of the exciton $|+\rangle$ is denoted by
$d_+$ and $f$ is a correction factor to the dipole moment in the
exciton-biexciton transition matrix element due to Coulomb
interaction.  The amplitude of the electric field $E_{+,j}(t-\tau_j)=
{\cal E}_{+,j}(t-\tau_j) e^{-i\omega_+(t-\tau_j)}e^{i \phi_j}$ is
assumed to be slowly varying.  As in the atomic case, the condition
$\omega_+ \gg d_+ {\cal E}_{+,j}$ enables the rotating wave
approximation used in $H^+$ above.  Thus, the counter-rotating terms,
such as $H^+_{0,-}=\Omega^*_+/2$, are set to zero. Similarly the
Hamiltonian associated with a $\sigma_-$ polarized electric field is
given by
\begin{equation} H^-=\left[\begin{array}{cccc}
0& 0 &\Omega_-(t)/2& 0 \\
 0& \epsilon_+& 0 & f \Omega_-(t)/2 \\
\Omega^*_-(t)/2& 0 & \epsilon_- & 0 \\ 
0& f \Omega^*_-(t)/2 & 0& \epsilon_{-+}
\end{array}\right]~.
\end{equation}

For simplicity of exposition we consider a sequence of non-overlapping
pulses although in numerical simulations we have found possible to
pack the pulses with 10\% overlap with negligible deterioration. Thus,
we write the unitary time evolution operator from $t=\tau_0$ to $t=T$
in the form
\begin{equation} 
U(T,\tau_0)=\prod_{j=1}^N
U^{\sigma_j}_j(\tau_{j},\tau_{j-1})~,
\label{product}
\end{equation} 
where $N$ indicates the number of pulses in the train and
$\tau_{j},\tau_{j-1}$ the beginning and the end of the $j$-th pulse.
For a given quantum operation $U(T,\tau_0)$, the time optimization can
be viewed as consisting of two components. The first is to have a
minimum number of pulses $N$ in Eq.~(\ref{product}).  Optical pulses
can directly perform Rabi rotations with generators $\sigma_x$ and
$\sigma_y$ but rotations with generator $\sigma_z$ need to be built as
a combination of $\sigma_y$ and $\sigma_x$.  In our design of the
laser implementation of a quantum algorithm we try to decompose the
required global transformation directly in rotations generated by
$\sigma_y$ and $\sigma_x$ for both single qubit and conditional
operations without appealing to Hadmard, C-NOT or conditional phase
shift.  We have demonstrated this by the construction of the
Deutsch-Josza algorithm\cite{chen} and the Quantum Fourier Transform
(see below). Since the saving is not exponential, in theory it may be
considered trivial but in practice, especially in the initial stage of
experimental implementation, the use of the right decomposition of the
algorithms may be advantageous.

The second component for a fast control is the time optimization of
each pulse in the product of Eq.~(\ref{product}), which is the main
subject of this section.  Consider the case of a $\sigma_+$ pulse. In
the interaction representation, $\tilde{O}=\Lambda O
\Lambda^{\dagger}$ denotes the transformed operator from $O$, with
$\Lambda(t)=e^{i H_0 t}$, where $H_0$ is a diagonal matrix with
elements $(0,\epsilon_+,\epsilon_-,\epsilon_{-+})$. The term
$U^{\sigma_j}_j$ in Eq.~(\ref{product}) becomes for $\sigma_+$ pulse
(with $j$ understood below)
\begin{equation}
\tilde{U}^{\sigma_+}= Te^{-i\frac{1}{2}\int_0^\tau dt
\tilde{V}^{\sigma_+}(t)}~,
\label{omega}
\end{equation}
 where $\tilde{V}(t)^{\sigma_+}$ is 
\begin{equation} 
\left[\begin{array}{cccc}
0& \Omega_{+}(t)e^{i\epsilon_{+}t} & 0 & 0 \\
\Omega^*_{+}(t)e^{-i\epsilon_{+}t}   & 0& 0 & 0 \\  
0 & 0 & 0 & f\Omega_{+}(t)e^{i(\epsilon_{+}-\Delta)t} \\  
0& 0& f\Omega^*_{+}(t)e^{-i(\epsilon_{+}-\Delta)t} & 0
\end{array}\right]\label{interact}
\end{equation}
and $\Delta=\epsilon_{+}+\epsilon_{-}-\epsilon_{-+}$ is the biexciton
binding energy.  When only a circularly polarized light is used,
Eq.~(\ref{interact}) shows that the four-level system behaves as a
double two-level system, the first two-level transition (exciton
transition) being represented by $|0\rangle \rightarrow |+\rangle$ and
the second (biexciton transition) by $|-\rangle \rightarrow
|-+\rangle$.

Consider now the desired operation where the exciton transition is a
Rabi rotation through angle $\alpha$ and the biexciton transition a
Rabi rotation through $\alpha^\prime$,
\begin{equation}
\tilde{U}^{\sigma_+}_j=\left[\begin{array}{cccc}
\cos(\alpha/2) & -\sin(\alpha/2) & 0 & 0 \\
\sin(\alpha/2) & \cos(\alpha/2) & 0 & 0 \\  
0 & 0 & \cos(\alpha^\prime/2) & -\sin(\alpha^\prime/2) \\  
0& 0& \sin(\alpha^\prime/2) & \cos(\alpha^\prime/2)
\end{array}\right].
\label{wantedcond}
\end{equation}
The most direct solution for the realization of this transformation
would be a two-pulse combination,
\begin{eqnarray}
E_{+}(t) &=& {\cal E}_0 e^{-(t/s)^2} e^{-i\omega_{0+}t} \nonumber \\
&+& {\cal E}_1 e^{-(t/s_1)^2}
e^{-i\omega_{1+}t+i\phi}.
\label{e2}
\end{eqnarray}
If the two pulses are resonant respectively with the two transitions,
i.e. $\omega_{0+}=\epsilon_+$ and
$\omega_{1+}=\epsilon_{-+}-\epsilon_-$, and sufficiently narrow in
frequency, the pulse resonant with the exciton transition would have
negligible effect on the biexciton transition and vice versa.
However, this has been shown to be costly in time.\cite{chen} The
problem is to find a composite pulse which would take much less time
with tolerable deterioration of quality of the transformation.

For the quality of the transformation, we follow Ref.~\onlinecite{poyatos} in
defining the fidelity of the transformation as
\begin{equation}
F=\overline{|\langle
\psi_{in}|\tilde{U}^\dagger U_i|\psi_{in}\rangle|^2},
\end{equation}

where $U_i$ is the ideal unitary operation, $\tilde{U}$ is the unitary
transformation generated by the optical pulses, and the overline
denotes the average over all the possible initial states. The operator
$\tilde{U}^\dagger U_i$ is denoted by $I$ for short. The average over
all the possible states is done by considering an initial state with
arbitrary complex coefficients $|\psi_{in}\rangle=\sum_j c_j|j\rangle$
with the normalization constraint $\sum_j |c_j|^2 =1$. The fidelity
can be then written in the form
\begin{equation}
F=\sum_{ijkl}\overline{c^*_i c_j c^*_k c_l} I_{ij} I^*_{lk}
\end{equation} 
and, in the four-level system considered here, the overline average is
then on a hypersphere S in $C^8$ determined by the normalization
condition. This average $(1/S)\int_S d^2c_1d^2c_2d^2c_3d^2c_4 c^*_i
c_j c^*_k c_l$ is easily evaluated in polar coordinates and gives
\begin{equation}
F=1/10
\sum_i |I_{ii}|^2 +1/20 \sum_{i\ne j} (I_{ii}I^*_{jj}+I^*_{ij}I_{ij}).
\end{equation}
The difference of the coefficients from those of
Ref.~\onlinecite{poyatos} is due to their additional restrictions on
the coefficients $c_j$. Our choice gives a more conservative
estimation of the error in the operations.

\subsection{ Pulse Design}

In this subsection, we explain three different approaches to pulse
design to shorten the time of the quantum operation.

\subsubsection{Approximation by the Area Theorem}

In the limit of very long pulses, the area theorem \cite{allen}
determines the intensity of a Gaussian pulse that has to be used for a
given rotation $\alpha$
\begin{equation} 
{\cal E}_0= \frac{\alpha}{s \sqrt{\pi} d_+}.
\label{area}
\end{equation}
For a single two-level system the pulse width $s$ in Eq.~(\ref{area})
can be made arbitrarily small, but in the four-level case we are
strongly limited by the resonance condition to $1/s,{\cal E}_0 d_+\ll
\Delta$. In order to shorten the time duration of the whole pulse, an
intuitive approach would be to allow the two components of
Eq.~(\ref{e2}) to overlap in frequency but keep each satisfying the
area theorem.

\subsubsection{The average Hamiltonian method}

The cumulant expansion (also known as the Magnus expansion
\cite{wilcox}) of the evolution operator $\tilde{U}^{\sigma_+}_j$ in
 Eq.~(\ref{omega}) is given by \cite{ernst87}
\begin{equation}
\tilde{U}^{\sigma_+}_j=e^{-\frac{i}{2} (\tilde{V}_1+\tilde{V}_2+\dots)}.
\label{magnus}
\end{equation}
The first term of the expansion corresponds to a time average of
the interaction Hamiltonian,
\begin{equation}
\tilde{V}_1=\int_{0}^{\infty}~dt \tilde{V}(t).
\end{equation}
The second term is given by
\begin{equation}
\tilde{V}_2=\frac{-i}{4}\int_{0}^{\infty}dt\int_{0}^{t}dt^\prime [
\tilde{V}(t), \tilde{V}(t^\prime)]. \label{v1}
\end{equation}
Keeping only the first term in the exponent constitutes the average Hamiltonian
approximation. An estimation of the error in the truncation of the cumulant
expansion is given by the second term.

\subsubsection{Numerical approach}

The parameters in Eq.~(\ref{e2}) are varied to find the maximum
fidelity. To lessen the numerical effort, physical considerations
guide the reduction of the number of parameters varied. The first two
approximation methods are also useful as starting points.

\subsection{Examples of pulse design}

We illustrate the above methods for a single qubit operation, i.e., a
parallel rotation of both the exciton and biexciton transitions. For
simplicity, let $f=1$ and $s_1=s$. Both theoretical estimates and
experimental measurements have the $f$ value not far from unity. In
any case, the extension to $f\ne 1$ can be made in a similar manner to
the treatment on the conditional rotation given below.  We consider a
composite pulse by superposing and phase-locking the the two pulses in
Eq.~(\ref{e2}) with ${\cal E}_0 = {\cal E}_1$ and
$\omega_{0+}=\epsilon_+$ and $\omega_{1+} =\epsilon_+ -\Delta$.  It
remains to choose a value for ${\cal E}_0(s)$ by each of the three
methods above and tests its efficacy by evaluating the fidelity of the
operation.

\begin{figure}
\centerline{\psfig{file=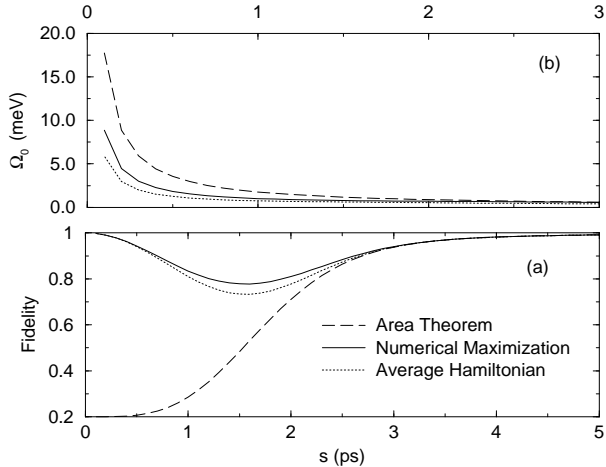,width=8. cm}}
\caption{ (a) Fidelity as a function of the temporal width of the
Gaussians $s$ for a parallel rotation of
$\alpha=\alpha^{\prime}=\pi$. (b) Peak value of the Rabi energy
$\Omega_0=d_+{\cal E}_0(s)$. Dashed lines: the area theorem
approximation.  Dotted lines: the average Hamiltonian approximation.
Solid Lines: numerical maximization of the fidelity. }
\label{fidelitypar}
\end{figure}

In Fig.~\ref{fidelitypar}(a) the fidelity for
$\alpha=\alpha^{\prime}=\pi$ rotation is plotted as a function of the
temporal width of the Gaussian pulse $s$. The corresponding value for
the peak of the Rabi energy $\Omega_0=d_+{\cal E}_0(s)$ is also given
in Fig.~\ref{fidelitypar}(b).  The value of the biexcitonic binding
energy $\Delta$ is 1 meV. The results by the area theorem
approximation are shown as the dashed lines. The fidelity is close to
unity only for $s \gg 1/\Delta$, corresponding to a region where the
frequency selectivity is preserved. If for instance a 98\% Fidelity is
required, the area theorem approach will lead to optical pulses with
$s> 4$ ps. The area theorem is not the best procedure of time
optimization for single-qubit operations.

Applying the average Hamiltonian approximation to the restricted pulse
specified above leads to the single-qubit rotation
$\tilde{U}^{\sigma_+}$ in the form of Eq.~(\ref{wantedcond}) with
chosen values for $\alpha=\alpha'$ and for $s$, leading to ${\cal E
}_0$ given by
\begin{equation}
{\cal E }_0=\frac{\alpha}{ s \sqrt{\pi} d_+ (1+e^{-(\Delta s/2)^2})}~.
\label{fourierpar}
\end{equation}
The Gaussian term in the denominator on the right gives a correction
to the area theorem, Eq.~(\ref{area}). The results are shown as dotted
lines in Fig.~\ref{fidelitypar}(b).  An estimate of the error of the
average Hamiltonian approximation may be made by evaluating the
second-order term in the cluster expansion given by Eq.~(\ref{v1}).  A
rough estimate is provided by replacing the Gaussians with square
pulses of width $s$,
\begin{eqnarray}
\tilde{V}_2 &=& (d_+{\cal E}_0/\Delta)^2 (\sin(\Delta s) -\Delta s \nonumber \\
&+& 2 \Delta s
\cos(\Delta s/2)-4\sin(\Delta s/2)) {\bf \Xi}= \phi{ \bf \Xi}~,
\end{eqnarray}
 where ${\bf \Xi}$ is a diagonal matrix with elements
$(-\frac{1}{2},\frac{1}{2},\frac{1}{2},-\frac{1}{2})$. Expanding in
the limit of short pulses $\Delta s \ll 1 $ we get $\phi \sim -
(d_+{\cal E }_0/\Delta)^2 (\Delta s)^3/3$. The correction to the area
theorem in the first order term, Eq.~(\ref{fourierpar}), is by
contrast $\sim (d_+{\cal E }_0/\Delta) (\Delta s)$.  Faster pulses
make the lowest order $\tilde{V}_1 \gg \tilde{V}_2$. The resultant
fidelity by the average Hamiltonian method is shown as dotted lines in
Fig.~\ref{fidelitypar}(a).  Note that it is possible to obtain a 98\%
Fidelity using much shorter pulses, of the order of 100 fs. In the
limit of very short pulses this correspond to pulses spectrally very
broad which do not distinguish between the two transitions but yield a
nearly parallel rotation.

The results of the numerical maximization using one variable ${\cal E
}_0$ by Brent's method \cite{nr} are plotted as solid lines.  The
optimal curve ${\cal E }_0(s)$ deviates considerably at short times
from the area theorem approximation but is close to the average
Hamiltonian approximation throughout the whole range of $s$.

The second example is a conditional operation for two qubits, viz., a
$\sigma_+$ biexcitonic transition without affecting the excitonic
$|+\rangle \rightarrow |0\rangle$, i.e., a rotation $\tilde{U}_j$ in
Eq.~(\ref{wantedcond}) with $\alpha=0$ and $\alpha^\prime=\pi$.  For
the combined pulse in Eq.~(\ref{e2}) we consider now $\phi=\pi$, and
again ${\cal E}_0 = {\cal E}_1$ and $\omega_{0+}=\epsilon_+,
\omega_{1+}= \epsilon_+ - \Delta$.

From the average Hamiltonian approximation (the first order term in
the cluster expansion), we obtain relations for the three parameters
of the pulse ${\cal E}_0$, $s$ and $s_1$ for the desired rotations,
\begin{eqnarray}
\alpha=d_+{\cal E }_0 \sqrt{\pi} (s-s_1e^{-(\Delta s_1/2)^2}),
\label{alpha0} \\
\alpha^\prime=d_+{\cal E }_0 \sqrt{\pi} (s_1-s e^{-(\Delta s/2)^2}).
\label{alphapi}
\end{eqnarray}
For a given value of $s_1$, the other two parameters may be solved in the 
case with $\alpha=0$ and $\alpha^\prime=\pi$,
\begin{eqnarray} s &= & s_1e^{-(\Delta s_1/2)^2}, \\
 {\cal E }_0 &=& \sqrt{\pi}/d_+(s_1-s e^{-(\Delta s/2)^2})~.
\end{eqnarray}
In the limit of large $\Delta$ the solution gives $s\rightarrow 0$
eliminating the term resonant with the excitonic transition and ${\cal
E}_0 \rightarrow \sqrt{\pi}/s_1d_+$ in accord with the area theorem
for the biexcitonic transition.  For $\Delta \ne 0$ this system has
always a solution for any $\alpha \ne \alpha^\prime$. Correction to
the average Hamiltonian approximation may be estimated in analogy to
the parallel rotation case in the limit $ s_1, s \ll 1/\Delta$ and
give for $\tilde{V}_2$ a diagonal matrix with elements
$(-\frac{\phi_1}{2},\frac{\phi_1}{2},-\frac{\phi_2}{2},\frac{\phi_2)}{2}$,
where
\begin{eqnarray}
\phi_1 &\sim & \frac{1}{32} (\frac{d_+{\cal E }_0}{\Delta})^2 [(\Delta s)^3+ 2
(\Delta s_1)^3-3 (\Delta s)(\Delta s_1)^2], \nonumber \\
\phi_2 &\sim& \frac{1}{96} (\frac{d_+{\cal E }_0}{\Delta})^2 [(\Delta
s_1)^3+ 2 (\Delta s)^3-3 (\Delta s_1)(\Delta s)^2]. \nonumber
\end{eqnarray}

\begin{figure}
\centerline{\psfig{file=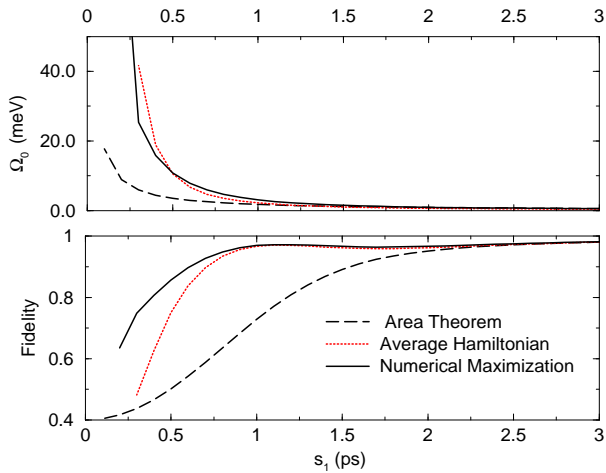,width=8. cm}}
\caption{(a) Fidelity as a function of the temporal width $s_1$ of the
biexciton Gaussian component in the composite pulse for a rotation of
$\alpha^\prime=\pi$ only for the biexciton transition. (b) Peak value
of the Rabi energy. Dashed lines: the area theorem approximation with
a single pulse resonant with the biexciton transition.  Dotted lines:
the averaged Hamiltonian approximation.  Solid Lines: numerical
maximization of the fidelity.}
\label{fidelitycrot}
\end{figure}

In Fig.~\ref{fidelitycrot} we show (a) the fidelity and (b) the peak
Rabi energy for the $\alpha=0$ and $\alpha^{\prime}=\pi$
transformation, for all three methods. The area theorem approximation
amounts to taking a single pulse resonant with the biexciton
transition. For the numerical maximization we maximize the fidelity
for a given $s_1$ value as a function of $s$ and ${\cal E}_0$ using
the downhill simplex method.\cite{nr} We see clearly that the average
Hamiltonian again gives a very good approximation: the deviations from
the numerical maximization are negligible in most of the region. Also
in this case we see that the use of a composite pulse provide a
considerable saving in the time for the operation.

\begin{figure}
\centerline{\psfig{file=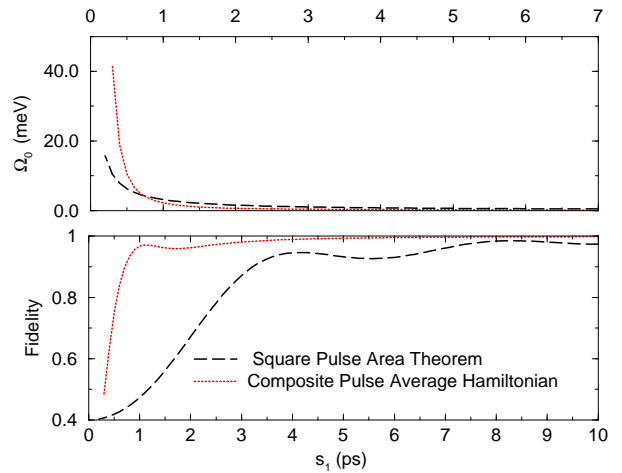,width=8. cm}}
\caption{(a) Fidelity as a function of the temporal width $s_1$ of the
square pulse and of the biexciton Gaussian component of the shaped
pulse for a rotation of $\alpha^\prime=\pi$ only for the biexciton
transition. (b) Peak value of the Rabi energy.  Dashed lines: when a
single square pulse resonant with the biexciton transition and the
area theorem is used.  Dotted lines: composite Gaussian pulse using
the average Hamiltonian approach, same as the dotted lines in
Fig.~\ref{fidelitycrot}.}
\label{square_pulse}
\end{figure}

As a last example, we investigate whether a single square pulse shape
can serve the function of the two overlapping pulses. For square pulse
an exact analytical expression for the $\tilde{U}^+$ can be given.  It
has been suggested \cite{berman} that off-resonant unwanted
transitions can be corrected using square pulses. In the specific case
discussed above this corresponds to the use of a single pulse resonant
with the biexciton transition with $s_1$ satisfying the conditions
\begin{eqnarray}
s_1 \Delta=\sqrt{4m^2-1} \pi \hbar~, \label{sp1}\\
s_1 \Omega_+= \pi~,\label{sp2}
\end{eqnarray}
with integer $m$. Eq.~(\ref{sp2}) gives a $\pi$ rotation for the
biexcitonic transition in accord with the area theorem, and the
condition in Eq.~(\ref{sp1}) sets to zero the off-diagonal terms in
the 2$\times$2 block corresponding to the excitonic
transition. However, additional phases in the diagonal corresponding
to a $\sigma_z$ rotations for the exciton transitions are introduced
which decrease the fidelity of the operation. We calculate the
fidelity and peak Rabi energy for a conditional $\pi$ rotation using a
single square pulse resonant with the biexcitonic transition as
functions of the temporal width $s_1$ of the square pulse and compare
it with the shaped pulse result of the average Hamiltonian
approximation in Fig.~\ref{square_pulse}. In the comparison, note that
$s_1$ in the square wave is the temporal width but in the shaped pulse
is the half width of the biexciton Gaussian component.  The fidelity
of the square wave shows oscillations with maxima roughly
corresponding to the conditions in Eqs.~(\ref{sp1}) and (\ref{sp2})
but never reaches as high as the two-pulse case.  Moreover, the spread
in frequency of the square pulse spectrum is a source for unintended
dynamics for higher exciton energy levels in the physical dot, while
Gaussian pulses avoid this problem.

\section{Quantum Fourier Transform}
\label{QFT}

The theory of control of the two excitons will now be applied to
construct a physical implementation of a quantum algorithm, the
two-qubit Quantum Fourier Transform. One qubit is given by the
presence or absence of a $\sigma^+$ polarized exciton in the dot and
the second qubit by the $\sigma^-$ exciton.  For the single-qubit
operation we have to act both in the exciton and in the biexciton
transition. The case of parallel $\pi$ rotation discussed in the
previous section corresponds, therefore, to a single-qubit operation
on the first qubit. As a conditional two-qubit gate we use a C-ROT
operation, which is essentially the C-NOT operation with a $\pi$
rotation replacing the logical NOT operation. The C-ROT is the
conditional dynamics of adding to the quantum dot a $\sigma_+$ exciton
only if an exciton with $\sigma_-$ polarization is already there. This
corresponds to a $\sigma_+$ biexcitonic transition\cite{barenco}
without affecting the excitonic $|+\rangle \rightarrow |0\rangle$.
The second example discussed in the previous section is a C-ROT
controlled by second qubit i.e.  by the $\sigma_-$ exciton.  Single
qubit and conditional rotations are easily generalized to arbitrary
angles. The exact mapping for the four-level system to two qubits is
given by
\begin{equation}
\{|0\rangle,|+\rangle,|-\rangle,|-+\rangle\}\rightarrow
\{|00\rangle,|01\rangle,|10\rangle,|11\rangle\}
\end{equation}
Unlike the NMR implementations, it is not possible here to make use of
the free evolution of the interacting qubits since it requires keeping
track of the oscillation at the optical frequency and, therefore, an
optical control over several picoseconds with sub-femtosecond
accuracy.  By working in the interaction representation, we get rid of
this drift term, making the design of the control more convenient.  At
the end of the sequence of pulses for a given algorithm the
interaction representation does not affect the computation since the
read-out is always done in a eigenstate of the system.  Therefore the
control of the qubit is always active and is constructed in terms of
rotations with $\sigma_x$ and $\sigma_y$ generators between pairs of
levels in the four-level system.

The Quantum Fourier Transform (QFT) is the key ingredient to a number
of important quantum algorithms, in particular Shor's
factorization. \cite{chuang00} Consider an $n$-qubit state
\begin{equation} 
|x\rangle \equiv |x_n \cdots x_1\rangle,
\mbox{where~} x=\sum_{i=1}^n x_i 2^{i-1}.
\end{equation}
The QFT is defined as a linear operator on an orthonormal basis of
$|0\rangle,\dots,|N-1\rangle$, where $N=2^n$, with the action
\begin{equation}
 U_{QFT}|x\rangle=\frac{1}{\sqrt{N}}\sum_{q=0}^{N-1}e^{2\pi
i x q/N}|q\rangle, \label{qftdef}
\end{equation}
analogous to the construction of Bloch states in a one dimensional
chain. The standard way to construct QFT employs two basic operations:
the Hadamard gate on the $j$th qubit $H_j$ and the conditional phase
gate $S_{jk}$, where $j$ is the control qubit and $k$ is the target
qubit.\cite{chuang00} The two-qubit QFT can be realized for instance
by the simple sequence $ H_2 S_{2,1} H_1$.  However, if we decompose
each of the three operations in Rabi rotations we end up in using more
optical pulses than necessary.  In fact, each Hadamard transformation
requires at least two optical pulses,
$R_j(\pi,\hat{x})R_j(\pi/2,\hat{y})$, where $R_j(\theta,\hat{e})$ the
rotation on $j$-th qubit in the $\hat{e}$ direction with angle
$\theta$.  Following the decomposition in
Ref.\onlinecite{weinstein01}, we find $S_{12}$ with the sequence,
\begin{eqnarray}
& R_1 (-\frac{\pi}{2},\hat{y}) CROT_{2,1}(\frac{\pi}{2},\hat{x})
 CROT_{\bar{2},1}(-\frac{\pi}{2},\hat{x})R_{1}(\frac{\pi}{2},\hat{x})
 & \nonumber \\ &
 R_{1}(\frac{\pi}{2},\hat{y})R_{2}(-\frac{\pi}{2},\hat{y})R_{2}(\frac{\pi}{2},\hat{x})
 R_{2}(\frac{\pi}{2},\hat{y})~. & \label{seqB}
\end{eqnarray}
In $CROT_{j,k}$, $j$ is the control qubit and $k$ is the target. For
example,
\begin{equation}
 CROT_{2,1}(\theta  ,\hat{x}) =\left[ \begin{array}{cccc}
 1 & 0 & 0 & 0 \\  0 & 1 & 0 & 0 \\
  0 & 0 & \cos(\theta/2) & -i\sin(\theta/2) \\  
0& 0& -i\sin(\theta/2) &
\cos(\theta/2)
\end{array}\right]~,
\end{equation}
and the bar over suffix 2 indicates a rotation on the target only for
the control qubit in the state 0,
\begin{equation}
 CROT_{\bar{2},1}(\theta  ,\hat{x}) =\left[ \begin{array}{cccc}
 \cos(\theta/2)  & -i\sin(\theta/2) & 0 & 0 \\  
  -i\sin(\theta/2) & \cos(\theta/2) & 0 & 0 \\
  0 & 0 & 1 & 0 \\  0& 0& 0 & 1
\end{array}\right]~.
\end{equation}
The total number of pulses for the QFT is then 12.

We redefine the QFT as
\begin{equation}  U_{MQFT}=B U_{QFT} \Sigma , \label{mqftdef}
\end{equation}
where $\Sigma$ is the all-qubit inversion ($x_i \rightarrow 1-x_i$),
\begin{equation}
\Sigma|x\rangle=|\bar{x}\rangle,
\mbox{where~} \bar{x}=\sum_{i=1}^n (1-x_i) 2^{i-1},
\end{equation}
and $B$ is the transformation ($x_i \rightarrow x_{n-i+1}$), which may be termed boustrophedon,\cite{bous}
\begin{equation} B|x_n \cdots x_2 x_1\rangle=|x_1 x_2 \cdots x_n\rangle=|\tilde{x}\rangle~.
\label{bou}
\end{equation}
In the Appendix, we prove that $U_{MQFT}$ is a composition of
rotations of generators $\sigma_x$ and $\sigma_y$ for states of any
number of qubits, denoted by $U_{MQFT}$. By avoiding the
pulse-consuming $S_{ij}$, this saves time by using a smaller number of
pulses than $U_{QFT}$.  $U_{MQFT}$ can be used directly in phase
estimation or factorization algorithms without the need for $B$ and
$\Sigma$, the global qubit transformations which are just re-labeling
of the qubits.  In a physical implementations there is the possibility
to make global qubit transformations that are simple relabeling, at no
cost from the point of view of the quantum control. If for instance a
quantum computer is composed of a chain of one-half spins, at any time
we can decide to flip all spin up in spin down and vice versa. This
all-bit inversion is a simple relabeling. We do not need to apply any
pulse to the chain; we have just to remember that in the readout. The
same can be done by switching in reading the string of qubits from the
right to the left instead than from the left to the right, which
corresponds to the boustrophedon transformation in Eq.~(\ref{bou}).
Although this saving in time is of the order polynomial in $n$, for
the current attempt at physical implementation of prototype quantum
computers it could provide a helpful simplification of the
experimental procedure.

For $n=2,N=4$ the pulse sequence for $U_{MQFT}$ is
\begin{equation} CROT_{2,1}(\frac{\pi}{2},\hat{x})R_2(-\frac{\pi}{2},\hat{y}) R_2(\frac{\pi}{4},\hat{x})R_1(-\frac{\pi}{2},\hat{y})~.
\label{sequence}
\end{equation}
 
We carried out a numerical simulation of the dynamics of the
multiexciton levels for this MQFT algorithm with and without the use
of composite pulses. We took the peak of the Rabi energy to be 2~meV,
larger than the 1~meV binding energy of the biexciton. The width of
the pulses is calculated using the area theorem approximation and the
average Hamiltonian. The corresponding values of fidelity for
$U_{MQFT}$ are 0.257 and 0.992. The pulse sequence is completed within
6 ps.

In order to check the robustness of the use of composite pulses in the
 presence of dephasing, we include the spontaneous emission in the
 simulation by adding the Lindblad operators in the equation of motion
 for the density matrix\cite{carmichael}
\begin{equation}
\frac{d}{dt}\rho= -\frac{i}{\hbar}[H,\rho] +\sum_{j=1}^4 \left(L_j \rho
L^\dagger_j
-\frac{1}{2}L^\dagger_j L_j \rho-\frac{1}{2}L^\dagger_j L_j
\right)~.
\end{equation}
where,
\begin{eqnarray}
L_1 = \sqrt{\Gamma}|0\rangle \langle +|, & & L_2=\sqrt{\Gamma}|0\rangle
\langle -|, \nonumber \\
L_3=\sqrt{\Gamma}|+\rangle \langle -+|, & &
L_4=\sqrt{\Gamma}|-\rangle \langle -+|,
\end{eqnarray}
 $\Gamma=15\mu eV$ being chosen to approximate the measured dephasing time.
\cite{bonadeoprl98}  These operators represent all the possible spontaneous
emission pathway in the four-level system.

There are many equivalent ways to solve the master equation in terms
of a nonlinear stochastic differential equation for a normalized state
vector $|\psi\rangle$.  We choose to use the quantum state diffusion
(QSD) equation \cite{gisin92}
\begin{eqnarray} |d\psi\rangle&=&-\frac{i}{\hbar}H|\psi\rangle dt \nonumber \\
&+&\sum_j \left( \langle L^\dagger_j\rangle L_j-\frac{1}{2}L^\dagger_j L_j
-\frac{1}{2}\langle L^\dagger_j \rangle \langle L_j \rangle
\right)|\psi\rangle dt
\nonumber\\ &+&\sum_j\left( L_j-\langle L_j \rangle \right) |\psi\rangle
d\eta_j
\end{eqnarray} 
where $\langle L \rangle=\langle \psi|L|\psi\rangle$ and
$\eta_j$ are independent complex random variables. The density matrix
can be expressed as $\rho=M|\psi\rangle\langle \psi|$ where $M$
denotes ensemble average and the expectation value of any operator $O$
is given by $M\langle \psi|O|\psi\rangle$. Inclusion of dephasing in
this way reduces the fidelity for the shaped pulse sequence of MQFT
from 0.992 to 0.892.

\section{Conclusions}
\label{discuss}

In the quantum control of multiexciton states in semiconductor quantum
dots, we have shown that the use of composite pulses makes possible
the realization of quantum operations in time scales of the order of a
hundred femtoseconds. In addition to the theory of methods of
constructing the pulses, we gave explicit examples to help
experimental implementation. We adopted the concept of fidelity as a
measure of the quality of a pulse sequence. We showed how to construct
a sequence of pulse based only on the physical $\sigma_x$ and
$\sigma_y$ rotations.  A numerical simulation of the application of
the shaped pulses to the two-qubit Quantum Fourier Transform in a
single semiconductor quantum dot provided a test of the pulse
shaping. While the work so far provides a complete blueprint for an
experimental demonstration of a simple quantum computation, future
work for a more realistic computer includes the inter-dot for scaling
up the system, design of optical control to minimize decoherence, and
design of optical implementation of quantum error corrections for
digital control of decoherence and unintended dynamics.

\acknowledgments This work was supported by ARO F0005010, NSF
DMR-0099572, and DARPA/ONR N0014-99-1-109. Y. S. D. acknowledges a
summer fellowship from California Institute for Telecommunications \&
Information Technology.

\onecolumn

\appendix \section{}
In this appendix the pulse sequences for the MQFT are constructed
for an arbitrary number of qubits. From Eqs.~(\ref{qftdef}) and
(\ref{mqftdef}),
the action of MQFT and the inverse are given by,
\begin{eqnarray}
 U_{MQFT}|x\rangle &=& \frac{1}{\sqrt{N}}
\sum_{q=0}^{N-1}e^{2\pi i \tilde{q}\bar{x}/N}|q\rangle, \\
U_{MQFT}^\dagger|q\rangle &=& \frac{1}{\sqrt{N}} =\sum_{x=0}^{N-1}
e^{-2\pi i \tilde{q}\bar{x}/N}|x\rangle.
\end{eqnarray}

Define $R_y$ to be a $y$-rotation on all the qubits. Then,
\begin{equation}
R_y|x\rangle \equiv \prod_j R_j(-\pi/2,\hat{y})|x\rangle
=\frac{1}{\sqrt{N}}
\sum_{p=0}^{N-1} e^{\sum_{j=1}^n \pi i p_j(1-x_j)}|p\rangle.
\end{equation}
 Now consider the combined transformation
\begin{eqnarray} R_yU_{MQFT}^\dagger|q\rangle&=&
\sum_{x=0}^{N-1} R_y |x\rangle \langle x|U_{MQFT}^\dagger|q\rangle=
\frac{1}{N}\sum_{p=0}^{N-1}\sum_{x=0}^{N-1}  e^{\sum_{j=1}^n \pi i p_j(1-x_j)}
e^{-2\pi\tilde{q}\bar{x}/N} |p\rangle \\ &=&
\frac{1}{N}\sum_{p=0}^{N-1}\prod_{j=1}^n
\sum_{x_j=0}^1 e^{\pi i (1-x_j)(p_j-\tilde{q}2^{j-n})} |p\rangle \\
&=& \sum_{p=0}^{N-1}\prod_{j=1}^n e^{-\pi i q^{(j)}/2}
\left(\cos(\pi q^{(j)}/2)\delta_{q_j,p_j}+i\sin(\pi
q^{(j)}/2)\delta_{q_j,1-p_j}
\right) |p\rangle \\
&=& \sum_{p=0}^{N-1} \prod_{k=2}^n e^{i\alpha_k q_k} \prod_{j=1}^n
\left(\cos(\pi q^{(j)}/2)\delta_{q_j,p_j}+i\sin(\pi
q^{(j)}/2)\delta_{q_j,1-p_j}
\right) |p\rangle \\ &=& \sum_{p=0}^{N-1} e^{i\alpha/2}
\prod_{k=2}^n \left(
e^{i\alpha_k/2}\delta_{q_k,1}+e^{-i\alpha_k/2}\delta_{q_k,0}
\right)\prod_{j=1}^n
\left(\cos(\pi q^{(j)}/2)\delta_{q_j,p_j}+i\sin(\pi
q^{(j)}/2)\delta_{q_j,1-p_j}
\right) |p\rangle. \label{last}
\end{eqnarray} where we use the definition $q^{(j)}=\sum_{k=j+1}^n q_k
2^{j-k}$.
The $\alpha_k$ is defined through the relation $\prod_{j=1}^n e^{-\pi
iq^{(j)}}=\prod_{k=2}^ne^{i\alpha_k q_k}$ and $\alpha=\sum_{k=2}^n
\alpha_k$. Note
that $\alpha_k$ depends only on $k$ and $N$.

\begin{figure}
\centerline{\psfig{file=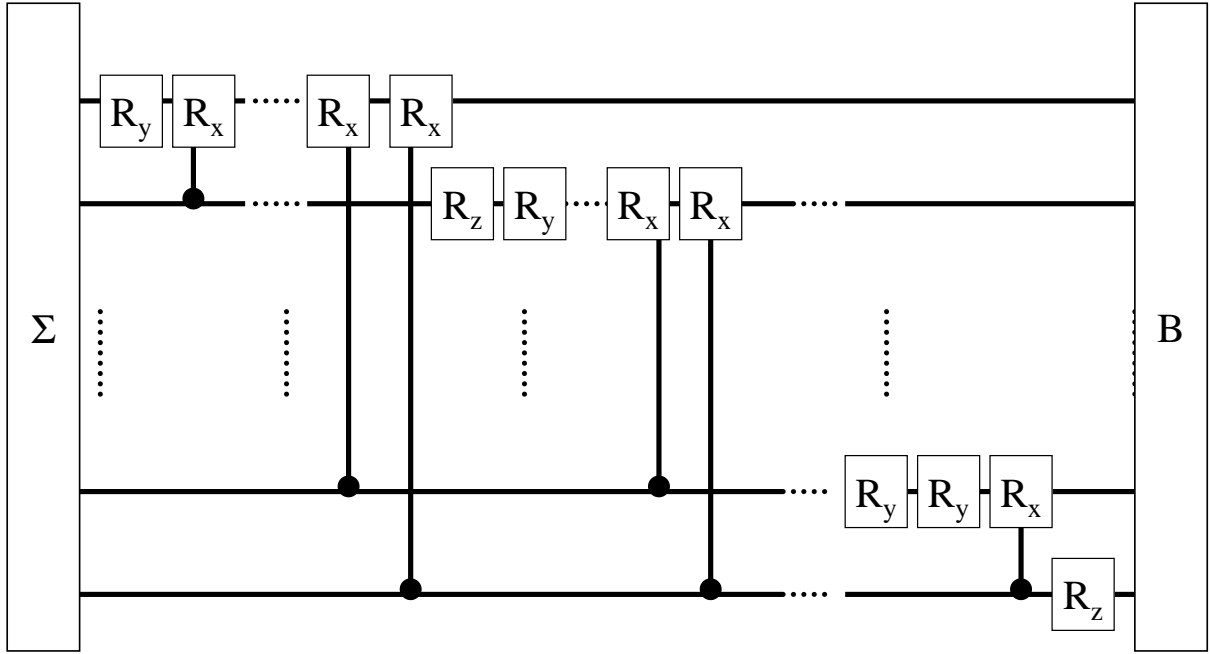,width=16. cm,angle=90}}
\vspace{3. cm}
\caption{Circuit diagram for QFT, Eq.~(\ref{A12}), with the operations
in the order from left to right. Each horizontal line represents a
qubit. The operations are explained in the text. The ones connecting
two quibit lines represent logic gates of controlled rotations.}
\label{qftscheme}
\end{figure}

The two products in the last line (\ref{last}) may be related to the rotations,
\begin{eqnarray}
\prod_{j=2}^n R_j(\alpha_k,\hat{z})|q\rangle &=& \prod_{j=2}^n
\left(e^{i\alpha_k/2}\delta_{q_k,1}+e^{-i\alpha_k/2}\delta_{q_k,0} \right)
|q\rangle, \\
 R_j(-\pi q^{(j)},\hat{x}) &=& R_j(-\pi\sum_{k=j+1}^n  q_k 2^{j-k},\hat{x})=
\prod_{k=j+1}^n CROT_{k,j}(-\pi 2^{j-k},\hat{x}).
\end{eqnarray}
These relations lead via
\begin{eqnarray} RU_{MQFT}^\dagger = e^{i\alpha/2}
\prod_{j=1}^n \prod_{k=j+1}^n CROT_{k,j}(-\pi 2^{j-k},\hat{x})
\prod_{k=2}^n R_k(\alpha_k,\hat{z}),
\end{eqnarray} to the conclusion that
\begin{eqnarray} U_{MQFT}=e^{-i\alpha/2} \prod_{j=1}^n \prod_{k=j+1}^n
CROT_{k,j}(\pi 2^{j-k},\hat{x})
\prod_{\ell=2}^n R_\ell(-\alpha_\ell,\hat{z}) \prod_{m=1}^n R_m(-\pi/2,\hat{y}). \label{A12}
\end{eqnarray}
Note that $CROT_{k,j}$ may be moved to the right as in the circuit diagram
Fig.~\ref{qftscheme} but not past any rotation involving the target qubit $j$.
Finally, by using
$R(\alpha,\hat{z})=R(-\pi/2,\hat{y})R(\alpha,\hat{x})R(\pi/2,\hat{y})$ we
obtain
a pulse sequence which involves only rotations and conditional rotations in the
$x,y$ direction to implement  MQFT. Note that the total number of
operations for
$n$ qubits is O($n^2$).

\vspace{3. cm}


\begin{references}
\bibitem{bous} Boustrophedon is writing that proceeds in one direction
in one line (such as from left to right) and then in the reverse
direction in the next line (such as from right to left).  The term
derives from the way one would plow land with an ox, turning the ox
back in the other direction at the end of a row. Boustrophedon is a
more efficient way to both write and read, especially if your lines
are very long.  Some types of printers and their software print in
this fashion.
\end{references}
\end{document}